\documentclass[letterpaper,12pt]{article}
\pdfoutput=1
\usepackage[letterpaper,margin=1in]{geometry}
\usepackage{graphicx}
\usepackage{amssymb}
\usepackage{siunitx}
\usepackage[colorlinks=true,
            urlcolor=blue,
            linkcolor=blue,
            citecolor=blue]{hyperref}
\usepackage{lineno}
\usepackage[T1]{fontenc}
\newcommand{\Mp}{m_\mathrm{P}}
\newcommand{\Lp}{\ell_\mathrm{P}}
\newcommand{\rS}{r_\mathrm{S}}
\begin{document} 

\begin{center}
\Large\textbf{%
Quasinormal modes of loop quantum black holes near the Planck scale
}
\end{center}

\centerline{Douglas M. Gingrich}

\begin{center}
\textit{%
Department of Physics, University of Alberta, Edmonton, AB T6G 2E1 Canada\\
\smallskip
TRIUMF, Vancouver, BC V6T 2A3 Canada
}
\end{center}

\begin{center}
e-mail:
\href{mailto:gingrich@ualberta.ca}{gingrich@ualberta.ca}
\end{center}

\centerline{\today}

\begin{abstract}
\noindent
We compare four loop quantum gravity inspired black hole metrics near
the Planck scale.
Spin 0, 1/2, 1, and 2 field perturbations on these backgrounds are
studied.
The axial-gravitational quasinormal modes are calculated and
compared. 
The time evolution of the ringdown is examined.
We also calculate the quasinormal modes in the eikonal limit and compare
with the predictions from circular null geodesics.
\end{abstract}
\section{Introduction}

First gravitational waves (GW) were detected in
2015~\cite{LIGOScientific:2016aoc, 
KAGRA:2022twx}.
The LIGO-Virgo-KAGRA Collaborations have since detected 90 GW-burst
events~\cite{KAGRA:2023pio}.
The Event Horizon Telescope revealed the first shadow images of
a supermassive object at the center of galaxy
Messier~87*~\cite{EventHorizonTelescope:2019dse} and
Sagittarius~A*~\cite{EventHorizonTelescope:2022wkp} in our
Milky Way. 
This observational data allows testing general relativity in the
strong-field regime.

The ability to deeply probe the gravitational domain has a bright
future with planned improved sensitivity and efficiency of existing GW 
detectors, as well as the proposed new detectors
LISA\footnote{https://lisa.nasa.gov/} and the Einstein
Telescope\footnote{https://www.et-gw.eu/}.  
There is a vision that future continuous improvements will enable
constraining new theories of gravity and possibly one day even shed
light on some quantum aspects of gravity.
Reconciling gravity with quantum mechanics can be considered one of
the grand challenges in theoretical physics.

Gravitational mergers have been observed and can be conceptualised to
occur in three stages: an initial inspiral, the merger, and a ringdown
stage.  
Insight into the ringdown stage can be described by perturbation
theory. 
A perturbed black hole is a dissipative system and gravitational waves
are emitted in a series of quasi-normal modes (QNM) which act as the
spectroscopy of the black hole. 
The amplitudes depend on source of the oscillations, while the
frequencies depend only on the black hole parameters. 
Since a proliferation of black hole merger observational data is
anticipated, it is important to study black hole QNMs.

The singularity in general relativity provides motivation to study
alternative theories of gravity.
It is anticipated that a complete quantum theory of gravity would be 
singular free.
Loop quantum gravity (LQG) has provided a paradigm in which to resolve
the singularity issue.
A number of black hole exterior semi-classical solutions have been
obtained under LQG~\cite{
Modesto:2008im,
Peltola:2008pa,
Bodendorfer:2019cyv, Bodendorfer:2019nvy, Bodendorfer:2019jay,
Ashtekar:2018lag,Ashtekar:2018cay,Ashtekar:2020ckv,
Gambini:2013ooa,Gambini:2013hna,Gambini:2020nsf,
Kelly:2020uwj,Kelly:2020lec,
Alonso-Bardaji:2022ear}.  
Each offering potentially interesting phenomenological implications.
LQG solves the singularity issue of the classical theory by
replacing it with an effective geometry with curvature of the order of 
the Planck scale or a transition surface.
The theory has a fundamental discreteness referred to as the area gap.
The effective solution is first obtained in a symmetry reduced phase
space framework and then expressed as a quantum corrected spacetime
metric.
Polymerization is used in the quantization procedure, where some
variables are replaced by their exponentiated versions. 

By studying the QNMs of LQG inspired black holes, we will be prepared
to confront the predictions with gravitational observational data.
Gathering information on different black holes solutions could help
elucidate the characteristics of alternative theories that satisfy
observational constraints, thus helping to validate, or discredit,
modified theories of gravity. 
In particular, it would be interesting to constraint the quantum
correction parameters of the black holes in LQG with the observed
gravitational wave ringdown signals.

The theme of this paper is to calculate and compare the eigenvalue
solutions due to perturbations on LQG black hole background metrics. 
The eikonal regime is also examined and compared with predictions
using circular null geodesics.
It is important to confront these prediction with LQG effective
solutions~\cite{Glampedakis:2019dqh}.
Most previous work on this topic study a single black hole for
different choices of its parameters.
In this work, we fix the parameters and compare the black holes for
the first time.

There is a rich literature on these topics in the context of modified
gravity under the LQG paradigm, thus showing their
importance~\cite{Modesto:2006gg, 
Modesto:2009ve,
Hossenfelder:2012tc,
Moulin:2019ekf,
Bouhmadi-Lopez:2020oia,
Daghigh:2020fmw,
Liu:2021djf,
Arbey:2021jif,
Albuquerque:2023lhm,
Fu:2023drp,
Moreira:2023cxy,
Yang:2023gas}.
Beside comparing the LQG cases among themselves, the standard
Schwarzschild solution is always presented alongside.
The comparisons with the metric by Modesto~\cite{Modesto:2008im} will
be different.
First, the phenomenology of this metric has been well
studied~\cite{Modesto:2006gg, Modesto:2009ve,
Hossenfelder:2012tc, Moulin:2019ekf, Arbey:2021jif, Yang:2023gas}.
Secondly, this metric has even been confronted with astronomical data
to set limits on its quantum parameters~\cite{Zhu:2020tcf,
Yan:2022fkr, Liu:2023vfh}. 
We use these constraints and thus the Modesto metric behaves
Schwarzschild-like.
It is not our business here to make value judgments of one black hole
solution over another, but at the end we cite controversies over one
of the metrics.
We mean no disrespect to the authors of many recent black hole
solutions not discussed here.  

When comparing different black holes, common geometry parameters need
to be chosen consistently. 
In addition, a common black hole characteristic depending on the mass,
such as the mass, horizon radius, Hawking temperature, etc.\ needs to
be chosen in the calculations. 
We will compare non-rotating and uncharged black holes, and so choose
a common Arnowitt-Deser-Misner (ADM) mass.  

Choosing a large mass corresponding to stellar, primordial, or
even small macroscopic black holes will lead to insignificant different
properties from Schwarzschild black holes for reasonable choices of
the parameters. 
To emphasize the unique features of the LQG black holes, we would like 
to choose the Planck mass $\Mp$ as the common mass value.
However, the LQG metrics are an effective solution as they only
approximate a full quantum theory of gravity.
The theory does not incorporate the full quantum gravity effects that
are needed near the Planck mass.
For macroscopic black hole masses $M \gtrsim 10^6 \Mp$, the effective
theories are valid.
LQG corrections that depend on the mass can become very small for
macroscopic black holes and are usually negligible for solar mass
black holes.
However, a full understanding of such quantum corrected black holes
requires investigation of features that are observable in principle if
not in practice. 
The results presented here will depend on the choice of black hole
mass.

For example, the potentials due to spin 0, 1/2, 1, and 2 perturbations
on the fixed black hole backgrounds are indistinguishable for the 
four different metrics we consider for $M \gtrsim 1000 \Mp$. 
For $M \approx 100 \Mp$, one of the potentials begins to decrease
slightly at the peak.
By $M \approx 10 \Mp$, a second potential starts to decrease at the 
peak.
Only when $M \approx 2 \Mp$, do three of the potentials deviate
from the Schwarzschild case.
One of the potentials remain identical to the Schwarzschild case even
when $M = \Mp$, as will be explained later.

When working near the Planck scale it is important to check any
approximations assuming large masses.
For calculational purposes, it will occasional be necessary to make
some approximations which will require a mass larger than the Planck
mass, yet can still be significantly below macroscopic black hole
masses. 

The layout of this paper is as follows.
In section~\ref{sec:2}, we write down the effective metric functions
and discuss their parameters.
The characteristic radii are discussed and the tortoise coordinate that
will be needed in some subsequent calculations is given.
Section~\ref{sec:3} discusses the choice for the parameters used in
the studies and the reasons for their choice.
In section~\ref{sec:5}, we write down the potentials due to different
spin-field perturbations on the background metrics.
The potentials will be the basis for subsequent calculations and
indicate the level of difference to anticipate in subsequent
results.
QNMs are calculated in section~\ref{sec:6}.
Ringdown and long-time tails are shown in section~\ref{sec:7}.
The eikonal limit is examined in section~\ref{sec:8}.
At the end of the paper the most important results are summarized and
discussed.

%
\section{Black hole metric functions and definitions\label{sec:2}}

In this section, we summarize the metric functions considered and
define their parameters.
We state the horizon radii and give the tortoise relationships.
The general static and spherically symmetric line element can be
written as 

\begin{equation}
ds^2 = -f(r)dt^2 + \frac{dr^2}{g(r)} + h(r) d\Omega^2.
\label{eq:line}
\end{equation}

In addition to the line element definition, we also define the general
tortoise relation between the tortoise coordinate $r_*$ and $r$: 

\begin{equation}
  \frac{dr_*}{dr} = \frac{1}{\sqrt{f(r)g(r)}}\, .
  \label{eq:tortoise}
\end{equation}

\noindent
This coordinate transformation will be essential to our calculation of
the quasi-normal modes.
It is one of the ingredients that allows us to transform the
radial equation of the perturbed metric to a wave equation with
short-range potentials. 
When solving Eq.~\ref{eq:tortoise} for $r_*$, the constant of
integration is not specified.
Fortunately, we more often need the differential equation, or it's
inverse, in the mathematics as opposed to $r_*$ itself.

Throughout, we work in geometric units of $G = c = 1$ and express
masses in units the Planck mass and distances in units of the Planck
length.
Also, when taking derivatives of metric functions, we use the continuous 
approximation.

\subsection{Kelly-Santacruz-Wilson-Erwing (KSW) metric}

Kelly, Santacruz, and Wilson-Ewing~\cite{Kelly:2020uwj} have obtained
a vacuum spherically symmetric spacetime using an effective framework
for the $\bar{\mu}$ scheme of holonomy corrections by imposing the
areal gauge in the classical theory and then expressing the remaining
components of the Ashtekar-Barbero connection in the Hamiltonian
constraint in terms of holonomies of the Planck length.
The resulting metric functions are

\begin{equation}
f(r) =  g(r) = 1 -\frac{\rS}{r} \left[ 1 - \frac{\gamma^2 \Delta
    \rS}{r^3} \right] \quad \textrm{and} \quad h(r) = r^2,
\end{equation}

\noindent
where $\rS = 2M$ and $M$ is the black hole ADM mass.
The parameters are the Barbero-Immirzi parameter $\gamma$ and the LQG
gap parameter $\Delta$ of the full theory.
This is the simplest metric we will examine as it involve only one
$1/r^4$ quantum correction term to $f(r)$ and $g(r)$, no correction to
$h(r)$, and effectively has only one free parameter, the combination 
$\gamma^2\Delta$.  

Two, one, or zero horizons are possible.
The horizon equation is quartic and we will use the numerical
solution.
To help aid understanding, expanding about $\gamma^2\Delta/\rS^2$
gives the following approximations. 
The outer horizon radius is approximately

\begin{equation}
r_+ = \left[ 1 - \frac{\gamma^2\Delta}{\rS^2}  - 3
\left( \frac{\gamma^2\Delta}{\rS^2} \right)^2 + \mathcal{O} \left(
\left( \frac{\gamma^2\Delta}{\rS^2} \right)^3 \right) \right] \rS\, . 
\end{equation}

\noindent
The inner horizon radius is approximately

\begin{equation}
r_- = \left[ \left( \frac{\gamma^2\Delta}{\rS^2} \right)^{1/3} +
\frac{1}{3} \left( \frac{\gamma^2\Delta}{\rS^2} \right)^{2/3}
+\frac{1}{3} \left( \frac{\gamma^2\Delta}{\rS^2} \right) + \mathcal{O}
\left( \left( \frac{\gamma^2\Delta}{\rS^2} \right)^{4/3} \right)
\right] \rS\, . 
\end{equation}

\noindent
A minimum radius exists, given by

\begin{equation}
r_\mathrm{min} = \left(\frac{\gamma^2\Delta}{\rS^2} \right)^{1/3} \rS\, .
\end{equation}

\noindent
A horizon radius exits, provided

\begin{equation}
M \ge \left( \frac{2}{\sqrt{3}} \right)^3 \sqrt{\gamma^2\Delta}.
\end{equation}

\noindent 
The use of the above metric functions is restricted by the
condition on the minimum mass and radius.

The tortoise relation is

\begin{equation}
\frac{dr_*}{dr} = \frac{1}{f(r)}.
\end{equation}

This form of the metric was also obtained more recently using a
different method~\cite{Lewandowski:2022zce}. 

\subsection{Gambini-Olmedo-Pullin (GOP) metric}

Gambini, Olmedo, and Pullin~\cite{Gambini:2020nsf} obtain an
interesting black hole metric based on improved
dynamics~\cite{Chiou:2012pg}.  
They include the ADM mass and its conjugate momentum in the scalar
constraint and represent the scalar constraint as an operator in the 
kinematical Hilbert space.
By applying group averaging techniques for both the quantum scalar
constraint and the group of finite spatial diffeomorphisms they obtain
the physical states, labeled by the ADM mass of the spin network.
Using some parameterized observables that act as local operators on
each vertex of the spin network they define the physical observables
denoting spacetime metric components. 
The resulting lowest-order metric is\footnote{We note a missing factor
of 2 in the cross term in Eq.~5.3 of \cite{Gambini:2020nsf}.} 

\begin{eqnarray}
f(r) & = & 1 - \frac{\rS}{r+r_0} + \frac{r_0^3 \rS^3}{(r+r_0)^6 (1 +
\frac{r_s}{r+r_0})^2},\nonumber\\
g(r) & = & f(r) \left[ 1 + \frac{\delta r}{2(r+r_0)} \right]^{-2},\\ 
h(r) & = & r^2\left[ 1 + \frac{r_0}{r} \right]^2 = r^2 +2r_0r +
r_0^2,\nonumber 
\end{eqnarray}

\noindent
where $\rS = 2M_0$ and

\begin{equation}
r_0 = \left( \frac{2M_0\Delta}{4\pi} \right)^{1/3}.
\end{equation}

\noindent
The $M_0$ parameter is related to the ADM mass by $M = M_0 + \delta
r/2$, 
where $\delta r$ is an independent quantization parameter.
The coordinate $r$ is discretized on a lattice with spacing $\delta r$
which typically would be chosen as the Planck length, $\delta{r} = \Lp$. 

The GOP authors have taken $\gamma = 1$ and their definition of the
gap parameter in terms of the KSW gap parameter is
$\Delta_\mathrm{GOP}(\gamma=1) \to
4\pi\gamma^2\Delta_\mathrm{KWS}(\gamma)$.
Thus GOP has the same minimum radius $r_0 = (\gamma^2\Delta
\rS)^{1/3}$ as KSW, accept $M_0 \ne M$ unless $\delta r\to 0$. 
The parameter $\delta r$ is a choice, but $r_0$ is dictated by $M$ and
$\delta r$. 

The horizon radius is given by solving a sixth-order polynomial of
which there is only one positive real root for our choice of
parameters, that is given approximately by 

\begin{equation}
r_\mathrm{h} = \left[ 1 -\frac{r_0}{\rS} -\frac{1}{4} \left(
  \frac{r_0}{\rS}\right)^3 + \mathcal{O} \left( \left( \frac{r_0}{\rS}
  \right)^6 \right) 
  \right] \rS\, . 
\end{equation}

\noindent
We will use the numerical solution in calculations.
The GOP horizon radius will be significantly smaller than the
Schwarzschild radius $2M$. 

The tortoise relation is

\begin{equation}
\frac{dr_*}{dr} = \left[ 1 + \frac{\delta r}{2(r+r_0)} \right] \frac{1}{f(r)}.
\end{equation}

\subsection{Ashtekar-Olmedo-Singh (AOS) metric}

In the Ashtekar, Olmedo, and Singh~\cite{Ashtekar:2018lag,
Ashtekar:2018cay, Ashtekar:2020ckv} approach, the plaquettes are
chosen to lie on the transition surface, that replaces the classical
singularity and separates the trapped region from the anti-trapped
region in the quantum extension of the Schwarzschild interior. 
The polymerization scales $\delta_b$ and $\delta_c$ are chosen to be
appropriate Dirac observables and a convenient choice of the laps
function determines explicit solutions to the effective dynamics.
The mass that characterizes the solution is a constant of the motion.
The effective spacetime structure is then extended to the exterior
region. 
The metric functions are

\begin{eqnarray}
f(r) & = & \frac{\left[ 2 + \epsilon + \epsilon
\left(\frac{\rS}{r}\right)^{1 + \epsilon} \right]^2}{\left[ 4 (1 + \epsilon)^2
\left( \frac{\rS}{r} \right)^\epsilon \right]^2}
\frac{  \left[ 1 - \left( \frac{\rS}{r} \right)^{1 + \epsilon} \right]
\left[ (2+\epsilon)^2 -\epsilon^2
 \left(\frac{\rS}{r}\right)^{1+\epsilon} \right]}{\left[ 1 + \left( \frac{L\rS}{r^2} \right)^2
\right] },\nonumber\\
g(r) & = & \frac{1}
{\left[ 2 + \epsilon + \epsilon \left(\frac{r_s}{r}\right)^{1+\epsilon}  
\right]^2}
\frac{\left[ 1 - \left(\frac{\rS}{r}\right)^{1+\epsilon}\right]
  \left[ (2+\epsilon)^2 - \epsilon^2 \left( \frac{\rS}{r}
  \right)^{1+\epsilon}
  \right] }{
  \left[ 1 + \left( \frac{L\rS}{r^2} \right)^2 \right]},\\ 
h(r) & = & r^2\left[ 1 + \left( \frac{L \rS}{r^2}\right)^2 \right]
= r^2 + \left( \frac{L \rS}{r}\right)^2,\nonumber
\end{eqnarray}

\noindent
where the polymerization scale $\delta_b$ is given by

\begin{equation}
  \epsilon = \sqrt{1+\gamma^2 \delta_b^2} - 1 \qquad \textrm{and} \qquad
  \delta_b = \left( \frac{\sqrt{\Delta}}{\sqrt{2\pi} \gamma^2 M}
    \right)^{1/3}\, . 
\end{equation}

\noindent
The polymerization scale $\delta_c$ is given by $L = \gamma L_0
\delta_c/4$, where 

\begin{equation}
L_0 \delta_c = \frac{1}{2} \left( \frac{\gamma\Delta^2}{4\pi^2M} 
\right)^{1/3},  
\end{equation}

\noindent
and $L_0$ is an infrared regulator.
None of the physical results depend on $L_0$ but on combinations with 
other parameters which are all independent of the choice of $L_0$.

These expressions for the effective metric functions are exact.
The effective metric approaches a flat metric as $1/r$ but the
curvature invariants fall-off slower than that of the Schwarzschild
metric. 
The ADM energy is well-defined and its value is $M$.

The outer event horizon is $\rS = 2M$ and the inner Cauchy horizon is
negligibly small $\sim \epsilon^2 \rS$.
The metric has two parameters which are fixed by LQG parameters and
the mass parameter.
The exponent parameter $\epsilon$ which decreases with mass could
have astrophysical consequences, while the length parameter $L$ is only
important near the Planck scale.

The tortoise relation is

\begin{equation}
\frac{dr_*}{dr} = \frac{ 4 (1+\epsilon)^2 \left(\frac{\rS}{r} \right)^\epsilon
\left[ 1 + \left( \frac{L \rS}{r^2} \right)^2 \right] }{ \left[ 1 - 
\left(\frac{\rS}{r}\right)^{1+\epsilon} \right] \left[ (2+\epsilon)^2
    - \epsilon^2\left( \frac{\rS}{r} \right)^{1+\epsilon}\right]}.
\end{equation}

\noindent
As $r\to \infty$,

\begin{equation}
\frac{dr_*}{dr} \to \left[ \frac{2(1+\epsilon)}{2+\epsilon} \right]^2 \left(
\frac{\rS}{r} \right)^\epsilon,
\end{equation}

\noindent
which will have important consequences later.

\subsection{Modesto metric}

The earliest loop quantum black hole metric we will consider is by
Modesto~\cite{Modesto:2008im}.
The spherically symmetric Hamiltonian constraint is modified in terms
of holonomies.
A large class of semiclassical solutions parameterized by a generic
function of the polymeric parameter are obtained.
The function choice is obtained by picking the one that reproduces
the Schwarzschild black hole solution outside the black hole with the
correct asymptotic flat limit.
The metric has been shown to have a $T$-duality and is often referred
to as the self-dual metric.
Of the metrics we consider, it has been the most extensively studied
and has a rich phenomenology~\cite{Modesto:2006gg, Modesto:2009ve,
Hossenfelder:2012tc, Arbey:2021jif, Moulin:2019ekf, Yang:2023gas}.
The metric functions are commonly written in the Reissner-Nordstr{\"o}m
form

\begin{eqnarray}
f(r) & = & \frac{(r-r_+)(r-r_-)}{r^4+a_0^2} (r+r_*)^2,\nonumber\\
g(r) & = & \frac{(r-r_+)(r-r_-)}{r^4+a_0^2} \frac{r^4}{(r+r_*)^2},\\
h(r) & = & r^2 \left[ 1 + \left( \frac{a_0}{r^2} \right)^2  \right]
= r^2 + \left( \frac{a_0}{r} \right)^2,\nonumber
\end{eqnarray}

\noindent
where $r_\mathrm{h} = r_+$ is the outer horizon radius, $r_-$ is a Cauchy inner
radius, and $r_* = \sqrt{r_+ r_-}$ (not the tortoise coordinate);
$a_0$ is related to the gap parameter in LQG by $a_0 = \Delta / (8\pi)$. 

The outer and inner horizons, and $r_*$ are given by

\begin{equation}
r_+ = \frac{1}{(1+P)^2} \rS,   \quad
r_- = \frac{P^2}{(1+P)^2} \rS, \quad
r_* = \frac{P}{(1+P)^2} \rS.
\end{equation}

\noindent
The polymeric function is

\begin{equation}
P = \frac{\sqrt{1+\epsilon^2}-1}{\sqrt{1+\epsilon^2}+1}, 
\end{equation}

\noindent
where $\epsilon = \gamma \delta$, with $\gamma$ the Barbero-Immirzi
parameter and $\delta$ the polymeric parameter.
The parameter $\delta$ and hence $P$ is a free choice.
The metric has two parameters which are independent of black hole mass.
The polymeric parameter could have astrophysical significance, while the
gap parameter is only important near the Planck scale.

The tortoise relation is

\begin{equation}
\frac{dr_*}{dr} = \frac{r^4+a_0^2}{(r-r_+)(r-r_-)r^2},
\end{equation}

\noindent
which can be analytically integrated to give

\begin{eqnarray}
r_* & = & r - \frac{a_0^2}{r_+r_-} \left[ \frac{1}{r} -
  \frac{r_++r_-}{r_+r_-} \ln(r) \right]\nonumber\\
& & + \frac{1}{r_+-r_-} \left[
  \frac{a_0^2+r_+^4}{r_+^2} \ln(r-r_+) - \frac{a_0^2+r_-^4}{r_-}
  \ln(r-r_-) \right]. 
\end{eqnarray}


\begin{center}
$\ast\ast\ast$
\end{center}

We now compare the metric functions and tortoise relations.
When comparing the metric functions $f(r)$ and $h(r)$ in different
back hole background metrics it would be beneficial to use some
standard coordinate system, such as, comparing $-g_{tt}$ after
transforming $h(r) \to r^2$ in each black hole background.
It is always possible to introduce such a new radial coordinate.
However, in this coordinate system the metric functions for some black
hole backgrounds (GOP, AOS, Modesto) become quite complicated
expressions and for the AOS and Modesto cases are double valued.
We thus continue to use the form of the metrics previously defined but 
keep in mind that the coordinate $r$ agrees only asymptotically with
the usual radial coordinate, and while for small $r$ it may bounce,
corresponding to a minimal possible area.

\begin{figure}[htb]
\centering
\includegraphics[width=0.49\linewidth]{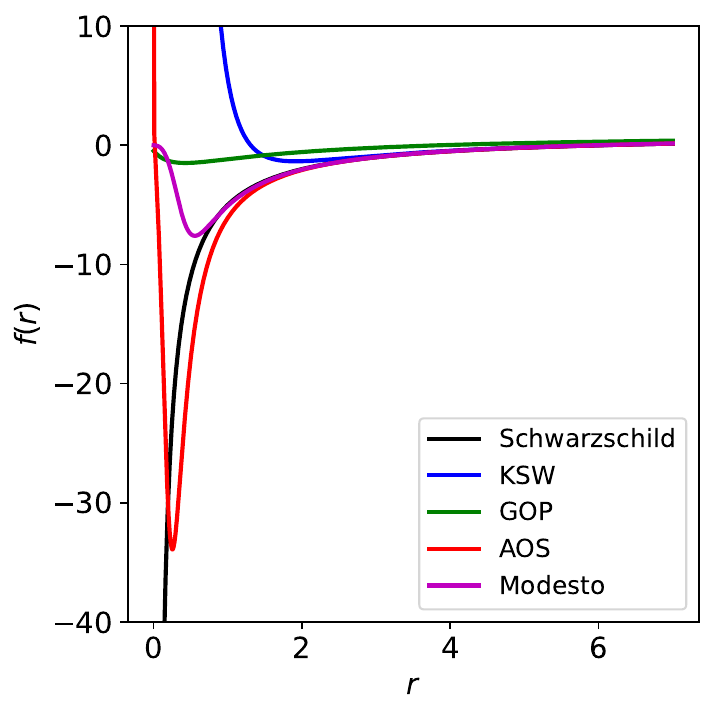}
\includegraphics[width=0.49\linewidth]{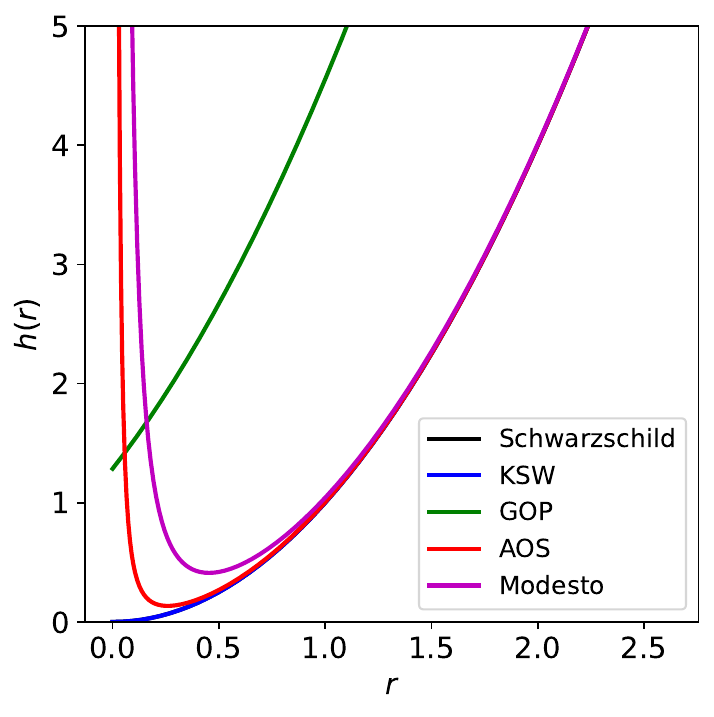}
\caption{\label{fig:metrics}(left) metric functions $f(r)$ and
(right) metric functions $h(r)$, for $M = 3 \Mp$.}
\end{figure}

Figure~\ref{fig:metrics}(left) shows the metric functions $f(r)$.
Outside the horizon only the GOP metric is noticeably different from
the others.
Only GOP has a single horizon.
Figure~\ref{fig:metrics}(right) shows the radial metric functions
$h(r)$.
AOS and Modesto show bounce.
GOP has a minimum $h(r) = r_0^2$ at $r=0$.
Except for GOP, $h(r)$ is negligibly different for $r \gtrsim \Lp$.
GOP $h(r)$ is significantly different over the $r$ we will consider.

\begin{figure}[htb]
\centering
\includegraphics[width=0.49\linewidth]{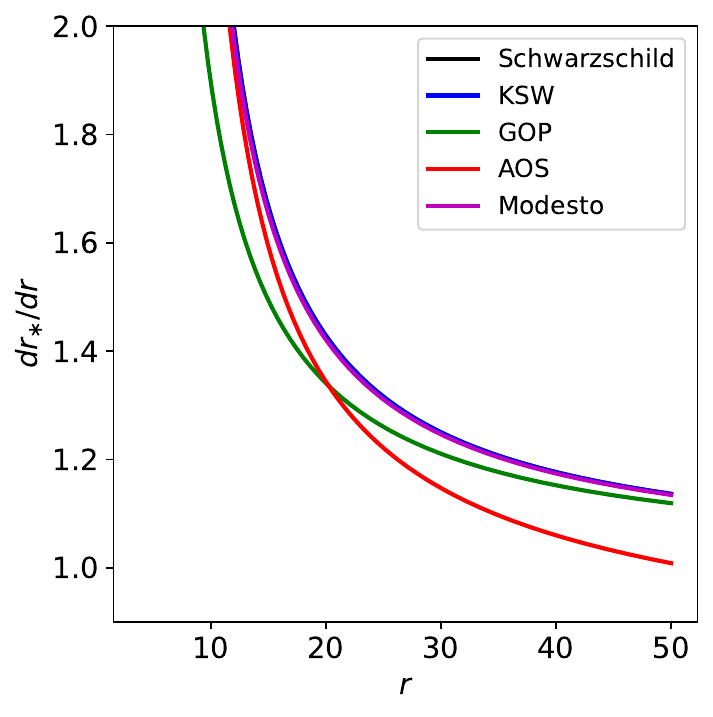}
\includegraphics[width=0.49\linewidth]{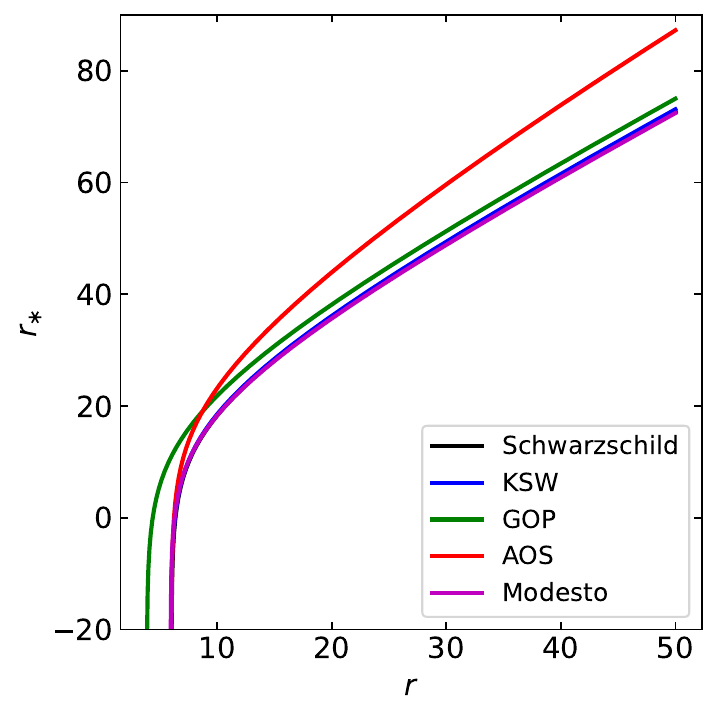}
\caption{\label{fig:rstar}(left) tortoise relations and (right)
tortoise coordinates, for $M = 3 \Mp$.}
\end{figure}

Figure~\ref{fig:rstar}(left) shows the tortoise coordinate
relationships.
Three different $dr_*/dr$ behaviors are observed.
GOP and AOS are mostly different due to $f(r) \ne g(r)$.

Analytic forms for the Schwarzschild, KSW, and Modesto tortoise
coordinate $r_*$ exist.
For the KSW, AOS, and GOP tortoise coordinate $r_*$ we, use the
analytic approximation 

\begin{equation}
r_* = \frac{1}{f^\prime(r_\mathrm{h})} \ln(r-r_\mathrm{h}),
+ \left. \int\frac{dr_*}{dr} \right|_{r\to\infty} dr,
\end{equation}

\noindent
where the constant has been set to zero.
The first term determines the behavior as $r\to r_\mathrm{h}$ and the second 
term as $r\to \infty$.
The second term integrates to $r$ for all metrics except AOS which
goes as $\sim r^{1-\epsilon}$.
This approximation does well near $r \to r_\mathrm{h}$ and $r \to \infty$.
Figure~\ref{fig:rstar}(right) shows the tortoise coordinate versus $r$.
The GOP $r_*$ is significantly shifted in $r$ due to its lower horizon
radius. 
The AOS $r_*$ is significantly different due to its large $r$ behavior.

%
\section{Choice of parameters\label{sec:3}}

In this section, we discuss our choice of parameters that are
subsequently used in calculations.
Some common LQG parameter values that we use are $\gamma = 0.2375$
for the Barbero-Immirzi parameter~\cite{Meissner:2004ju} and $\Delta
= 4\pi\sqrt{3}\gamma\Lp^2 = 5.169\Lp^2$ for the gap parameter.
All the metrics contain these parameters so we use common values.

The KSW parameter combination is $\gamma^2\Delta =
4\pi\sqrt{3}\gamma^3\Lp^2 = 0.292\Lp^2$. 
The minimum mass to have a horizon radius is $M \ge 0.8 \Mp$, which we
satisfy. 

For the GOP parameters, if $M$ and $\delta r$ are in Planck units,

\begin{equation}
r_0 = \gamma \left[ 8\pi\sqrt{3} \left( M - \frac{\delta
r}{2} \right)  \right]^{1/3} \Lp = 0.8355  
\left( M - \frac{\delta r}{2} \right) ^{1/3}\Lp\, . 
\end{equation}

\noindent
The restriction $\Lp^2/(2r_0) \le \delta r\le
r_0$~\cite{Gambini:2020nsf} means $M \gtrsim 2.1\Mp$. 
Assuming $\delta r \ge \Lp$, we will take $\delta r = \Lp$.
Hence we pick $M = 3\Mp$ as our working mass based on the free choice
of the discretization parameter $\delta r$.
For $M = 3\Mp$, $r_0 = 1.134\Lp$.

For the AOS parameters, 

\begin{equation}
\gamma^2\delta_b^2 = \left( \frac{\gamma^2\Delta}{2\pi
M^2} \right)^{1/3} =
(2\sqrt{3})^{1/3} \gamma \left( \frac{\Mp}{M} \right)^{2/3}\, . 
\end{equation}

\noindent
For $M = 3\Mp$, $\gamma^2\delta_b^2 = (2\sqrt{3})^{1/3}\gamma$ and
$\epsilon  = 0.0829$.
Also,

\begin{equation}
L = \frac{1}{8} \left(\frac{12}{M}\right)^{1/3} \gamma^2 \Lp\, .
\end{equation}

\noindent
For $M = 3\Mp$, $L = \frac{12^{1/3}}{8} \gamma^2\Lp = 0.0112\Lp$.

For the Modesto polymeric function, we consider constraints from
data~\cite{Zhu:2020tcf,Yan:2022fkr,Liu:2023vfh}. 
Since $P$ is expected to be a universal parameter for all black hole
masses these limits are applicable to our case of a mass near the
Planck scale.
The most recent upper limit is  $P < 6.17\times 10^{-3}$
on the polymeric function and $|\delta| < 0.67$ on the polymeric
parameter, at the 95\% confidence level~\cite{Liu:2023vfh}.
We take this limit as our parameter value for $P$.
The minimum area parameter is $a_0 = \Delta/(8\pi)
= \sqrt{3}/2\gamma\Lp^2 = 0.206\Lp^2$. 

All the parameters used in this paper are summarized in
Table~\ref{parm}.
The parameters $\gamma$ and $\Delta$ are fixed.
$\delta r$ is a choice, while $P$ is based on constraints to data.
The remaining parameters are fixed by the choice of the black hole
mass.

\begin{figure}[htb]
\centering
\includegraphics[width=0.6\linewidth]{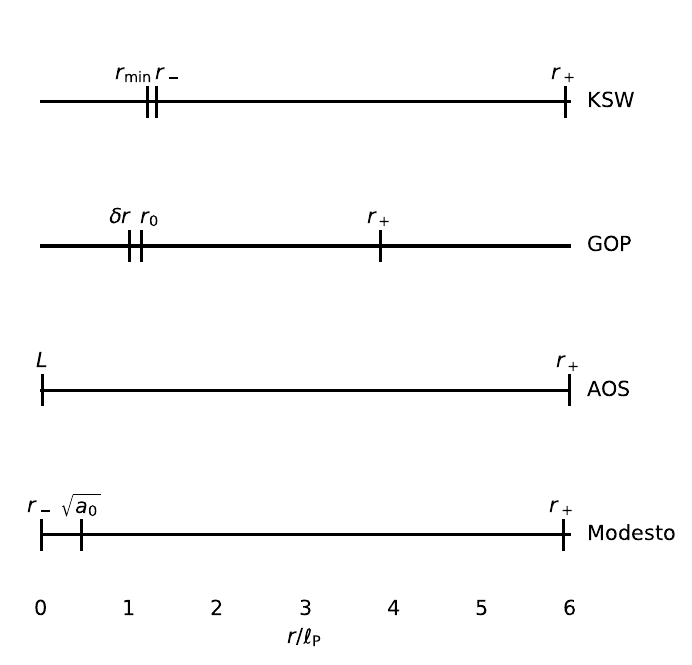}
\caption{\label{fig:scales}Distance scales for $M = 3 \Mp$.}
\end{figure}

The horizon radii using these parameters are shown in
Table~\ref{tab:radii}.
Also shown in Table~\ref{tab:radii} are radii of circular null
geodesics $r_c$ and shadow radii $R_\mathrm{s}$, which we
will discuss late. 
Figure~\ref{fig:scales} shows the distance scales for $M = 3\Mp$.
We notice that $L$ of AOS and $r_-$ of Modesto are negligible.
The two GOP parameters $r_0$ and $\delta r$ are both significant.
The Modesto parameter $a_0$ is not particularly significant.

\begin{table}[htb]
\centering
\caption{Parameters values. If the parameter depends on mass, $M = 3 \Mp$
has been taken.\label{parm}}
\vspace{0.2cm}
\begin{tabular}{ll}\hline
Parameter & Value\\\hline\hline
$\gamma^2\Delta$ & $0.292\Lp^2$\\\hline
$r_0$            & $1.134\Lp$\\
$\delta r$       & $\Lp$\\\hline
$\epsilon$       & $0.0829$\\
$L$              & $0.0112\Lp$\\\hline
$P$              & $0.00617$\\ 
$a_0$            & $0.206\Lp^2$\\
\hline
\end{tabular}
\end{table}

\begin{table}[htb]
\centering
\caption{\label{tab:radii}Characteristic radii in units of $\Lp$ for $M =  3\Mp$:
$r_\mathrm{h}$ is the outer horizon radius,
$r_c$ is the circular null geodesic radius, and
$R_\mathrm{s}$ is the shadow radius.}
\vspace{0.2cm}
\begin{tabular}{lccc}\hline
& $r_\mathrm{h}$ & $r_c$ & $R_\mathrm{s}$\\\hline
Schwarzschild & 6    & 9    & 15.59\\
KSW           & 5.95 & 8.96 & 15.55\\
GOP           & 3.85 & 6.36 & 12.98\\
AOS           & 6    & 9.12 & 15.39\\
Modesto       & 5.93 & 8.88 & 15.33\\
\hline
\end{tabular}
\end{table}

%
\section{Potentials\label{sec:5}}

External perturbations of black holes can be represented by
incident waves of different spins.
We adapt the test-particle point of view in which we consider the
propagation of waves in the black hole spacetime.
The perturbations due to a massless field of spin $s$ on a black hole
background spacetime have been well studied~\cite{Chandrasekhar}.
Different methods of solution are possible and the results depend on
both the spin of the perturbing field and the background spacetime. 
For the scalar field it is common to solve the Klein-Gordon equation
in curved space time and for the other spins to employ the
Newman-Penrose formalism~\cite{10.1063/1.1724257}.
For the general line element of Eq.~\ref{eq:line}, it is possible to
obtain a wave equation for each spin field.
The general potential in the wave equation thus only needs to be
written in terms of each black hole's metric functions, rather than
performing separate derivations for each metric.

For gravitational perturbations, it is common to either linearlize
the Einstein equations or to apply the Newman-Penrose
formalism~\cite{10.1063/1.1724257}. 
It is usual to consider the axial-gravitational perturbations
separately from the polar-perturbations.
They both give rise to wave equations but with different potentials.
For the case of Schwarzschild black holes, and some others, the
resulting spectra from these two different potentials are identical,
or so called, isospectral~\cite{Chandrasekhar}.  
Isospectrality is not guaranteed for all
metrics~\cite{Moulin:2019bfh}. 
Both methods of solution have been applied to the general metric
Eq.~\ref{eq:line} to obtain the axial perturbations.
Thus for each metric considered here, we only need calculate the
potential using the different metric functions, and do not repeat the
derivations.

The perturbation due to a field of spin $s$ on a black hole background
spacetime can be written in the Schr{\"o}dinger-like form as

\begin{equation}
\left[ \frac{d^2}{d r_*^2} + \omega^2 - V_s(r) \right] \psi(r_*) = 0,
\label{eq:schr}
\end{equation}

\noindent
where $\omega$ is the quasinormal mode frequency and $r_*$ is the
tortoise coordinate. 
The effective potentials for spin 0, 1/2, 1, and 2 can be written
as~\cite{Arbey:2021jif} 

\begin{eqnarray}
V_0(r)     & = & \nu_0     \frac{f}{h} +
\frac{\partial_*^2\sqrt{h}}{\sqrt{h}},\label{eq:V0}\\ 
V_{1/2}(r) & = & \nu_{1/2} \frac{f}{h } \pm \sqrt{\nu_{12}} \partial_*
                           \sqrt{\frac{f}{h}},\\
V_1(r)     & = & \nu_1     \frac{f}{h},\\ 
V_2(r)     & = & \nu_2     \frac{f}{h} + \frac{(\partial_* h)^2}{2h^2} -
                           \frac{\partial_*^2\sqrt{h}}{\sqrt{h}},\label{eq:axial}
\end{eqnarray}

\noindent
where
$\nu_0 = \nu_1= \ell(\ell+1)$,
$\nu_{1/2} = \ell(\ell+1) + 1/4$, 
$\nu_2 = \ell(\ell+1)-2$, and $\ell$ is the multipole number. 
The metric functions $f$ and $h$ are functions of $r$, and the
derivatives are with respect to $r_*$.
The first term in the different potentials is leading in $1/r$
behaviour to within an $\ell$-dependent constant. 
The $\ell(\ell+1)$ piece of the first term is the ``centrifugal
term''.

\begin{figure}[p]
\centering
\includegraphics[width=\linewidth]{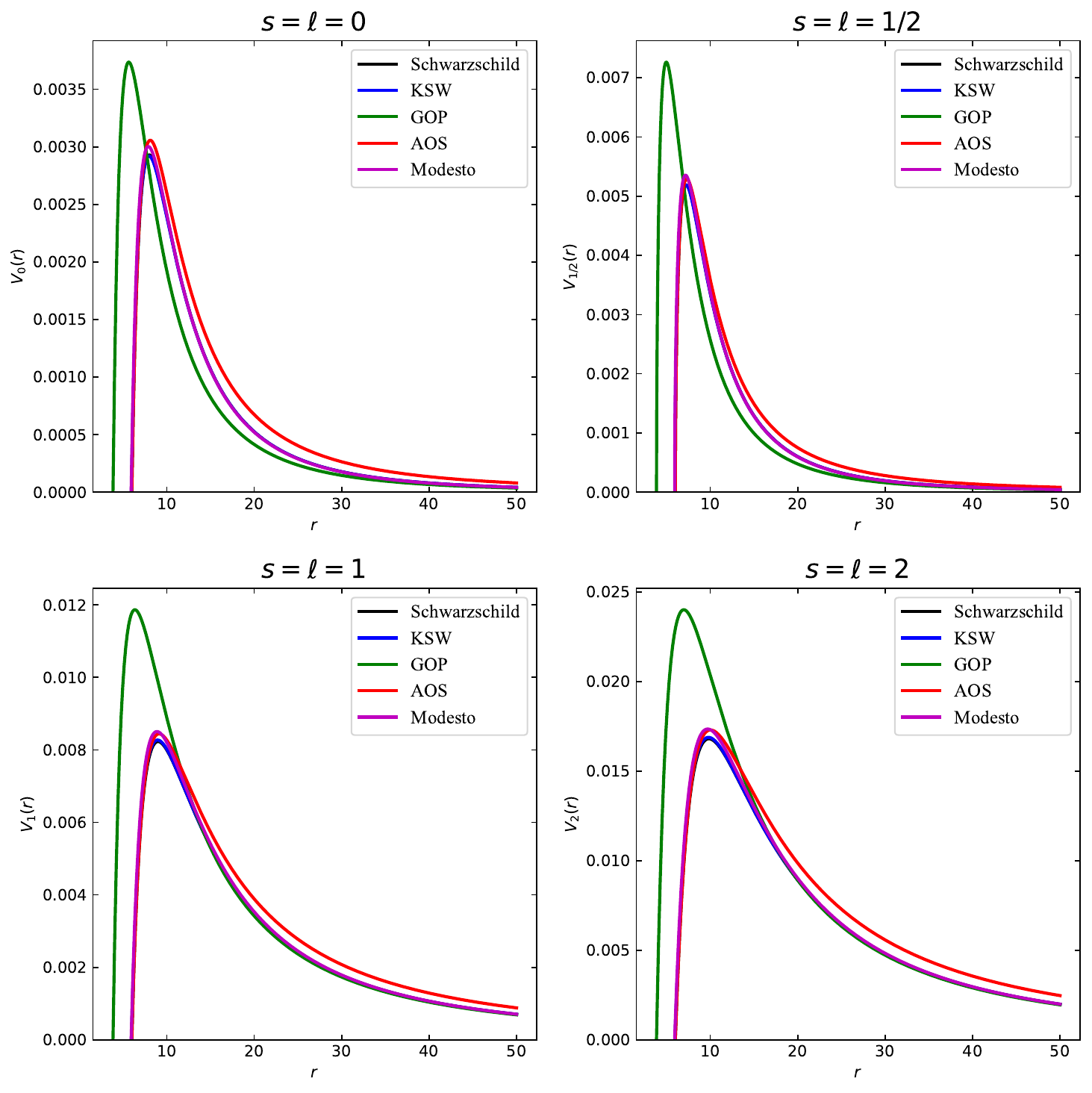}
\caption{\label{fig:pot}Potentials versus $r$ for spin  0, 1/2, 1, and
2 perturbations on the loop quantum gravity black hole background
metrics. 
$M = 3\Mp$ has been used.
In some cases, the Schwarzschild curve is hidden under the others.}
\end{figure}

\begin{figure}[p]
\centering
\includegraphics[width=\linewidth]{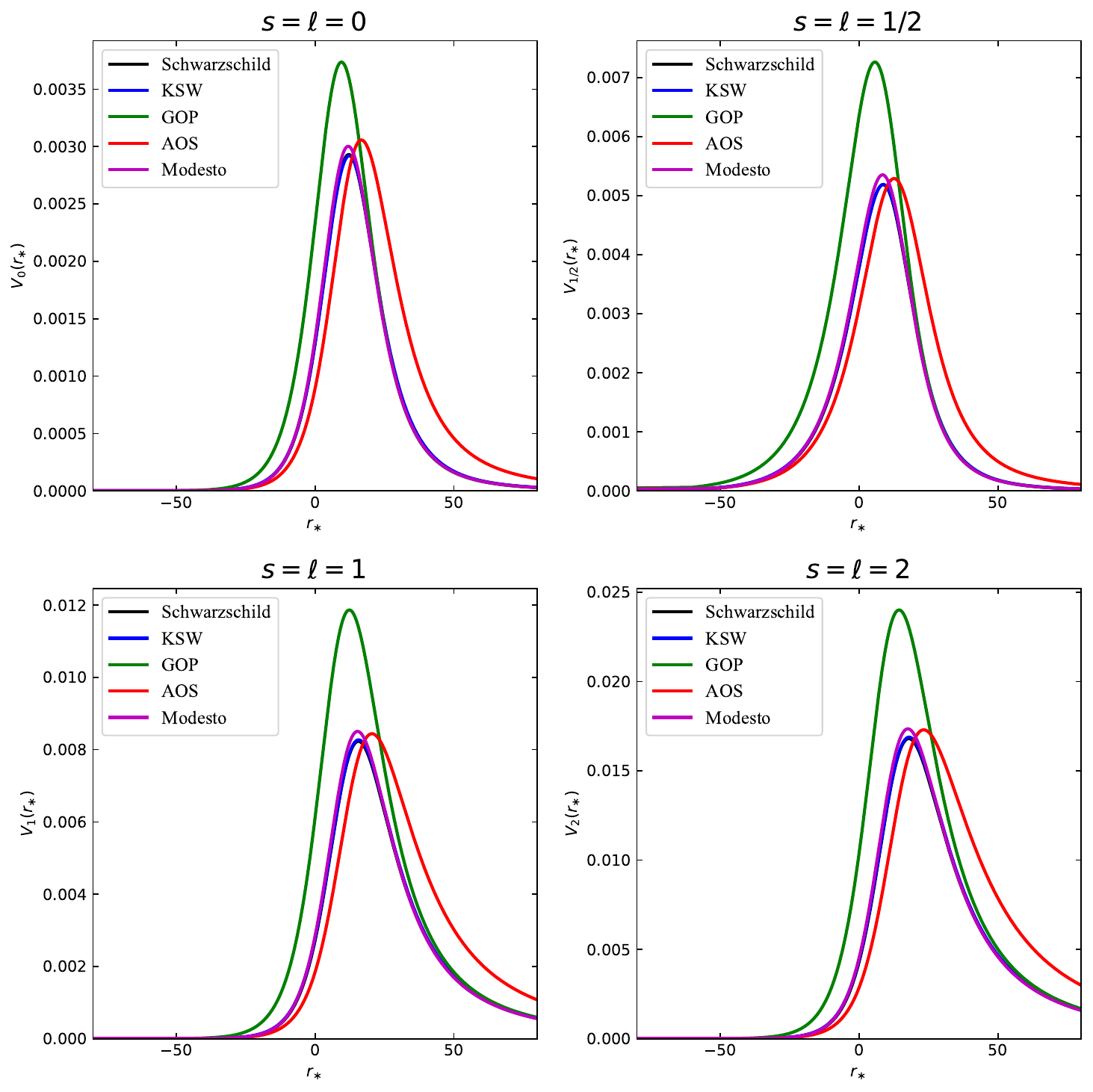}
\caption{\label{fig:pots}Potentials versus $r_*$ for spin  0, 1/2, 1, and
2 perturbations on the loop quantum gravity black hole background metrics.
$M = 3\Mp$ has been used.
In some cases, the Schwarzschild curve is hidden under the others.}
\end{figure}

Figure~\ref{fig:pot} and \ref{fig:pots} show the potentials versus
$r$ and versus $r_*$, respectively.
The potentials do not depend on the metric function $g(r)$ directly,
but only through the tortoise coordinate transformation.
The GOP potential is noticeably different; it begins to rise earlier
and has a higher maximum.
Both these differences are due to the smaller horizon radius.
The lower range of the potential is given by the horizon radius and
the maximum goes as approximately $1/r_\mathrm{h}$.
For black holes with similar horizon radii, the differences in the
metric functions outside the horizon is not great enough to show
significantly different potentials for most spins. 
However, for the $s=\ell=0$ case, the $1/r$ term in the potential
vanishes and the dependence on the tortoise relationship in the second
term of Eq.~\ref{eq:V0} is significant enough to change the
potential from the Schwarzschild case for the GOP and AOS black holes. 
In Fig.~\ref{fig:pots}, the differences for GOP and AOS are mainly due
to the difference in the tortoise relationship shown in
Fig.~\ref{fig:rstar} compared to the Schwarzschild case. 

As mentioned previously, the gravitational perturbations of a
spherically symmetric black hole can be divided into odd (axial)
parity and even (polar) parity under space inversion.
In what follows, we only consider the axial-symmetric gravitational
perturbations.
Moreover, since the effective potentials are everywhere positive
definite, we expect the black holes to be stable against all spin
perturbations. 

%
\section{Quasinormal modes\label{sec:6}}

In this section, we compute the axial-gravitational QNMs.
Given the Schr{\"o}dinger-like wave Eq.~\ref{eq:schr}, and the
potential due to axial-gravitational perturbations
Eq.~\ref{eq:axial}, the problem becomes an eigenvalue problem for
the complex frequencies $\omega$.

There are many methods that can be used to solve for the eigenvalues
due to the perturbations.
We employ the Wentzel-Kramers-Brillouin (WKB) approximation
method~\cite{1985ApJ...291L..33S, will1986approximation}.
The method is semi-analytic and only requires the derivatives of
$\omega^2 - V_s(r(r_*))$ evaluated at the maximum in $r_*$.
A Taylor series expansion of the potential about the maximum is used
with increasing number of terms leading to higher orders of
approximation. 

The WKB method uses the following generic formula to order $p$:

\begin{equation}
\omega^2 = V_0 - i \sqrt{-2 V_0^{\prime\prime}} \left[ n + \frac{1}{2}
  + \sum_{i=2}^{p}\Lambda_i \right], \quad n=0,1,\ldots\, ,
\end{equation}

\noindent
where $V_0^{\prime\prime}$ is the second derivative of $V(r_*)$
calculated at the maximum of the potential, and $n$ is the overtone
number. 
For example, $n=0, \ell = 2$ is the fundamental quadrapolar mode.
The method depends only on $\ell$ through the potential.
The correction terms $\Lambda_i$ represent the higher orders of 
approximation and depend on higher-order derivatives of the potential. 

The WKB method is valid only for asymptotically flat spacetime, or
for spacetime admitting wavelike solutions at spatial infinity and is
not applicable for $\ell < n$.
The WKB formula converges only asymptotically and does not guarantee
convergence in each order.
The method is not particularly accurate for low $\ell$ and high $n$.
We employ a third-order WKB method beyond the eikonal
approximation~\cite{PhysRevD.35.3621}, which uses up to sixth-order
derivatives in the correction terms.
Higher-order methods exist \cite{PhysRevD.68.024018,Konoplya:2019hlu}.
Although many QNM studies use multiple methods, that is not our intent
here. 
For qualitative comparison purposes, the third-order WKB approximation
is adequate.

%
%
%
%

For the GOP, AOS, and Modesto potentials, the number of terms
in the sixth derivative is beyond our computational abilities.
For the GOP case, we make a change of variable $r + r_0 \to r$,
approximate $f(r)$ to order $1/r^4$, and expand the Jacobian
$dr/dr_*$ to first order in $\delta r/r$, 
which is good to better than 1\%.
For the AOS case, we use the improved small $\epsilon$ approximation
of Ref.~\cite{Daghigh:2020fmw} which is good to order $\epsilon$.
For the Modesto case, we approximate $r_-/r \to 0$.
These are the only analytical approximations made in this paper and
are only made when calculating the derivatives for the WKB method.

\begin{figure}[htb]
\centering
\includegraphics[width=0.6\linewidth]{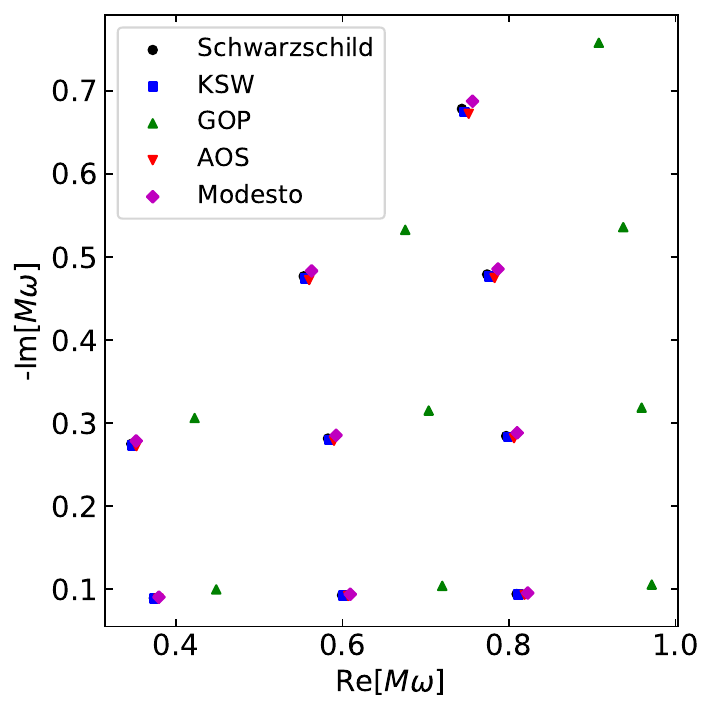}
\caption{\label{fig:quasi}Axial-gravity quasinormal modes for loop quantum
gravity black holes for $\ell = 2, 3, 4$ and $0 \le n < \ell$.
The points for a given metric have $\ell$ values that increase from
left to right and $n$ values that increase from bottom to top.}
\end{figure}

The QNMs are shown in Fig.~\ref{fig:quasi}.
The blocks of points are $\ell = 2$, $\ell = 3$, and $\ell = 4$ when
going from left to right, while $n$ increases from bottom to top as $0
\le n < \ell$. 
We reproduce the Schwarzschild QNMs for all modes examined, and
agree well with previous results for the other cases where available.
The GOP values agree qualitatively with Ref.~\cite{Liu:2021djf}.
That paper looks at massless scalar, electromagnetic,
axial-gravitational perturbations. 
The paper uses $M_0 = 10$ and $\delta r$ very small, which makes
quantitative comparisons difficult.
The AOS values agree qualitatively with Ref.~\cite{Daghigh:2020fmw},
which considers scalar perturbations.
GOP and AOS black holes oscillate with higher frequency and less
damping than Schwarzschild black holes. 
We found no modes with positive damping, which indicates these LQG black
holes are stable against massless axial-gravitational perturbations
that we examined. 

%

\section{Ringdown waveforms\label{sec:7}}

In this section, we calculate the time-domain profile of an
axial-gravitational perturbation on the background given by the
black hole metrics.
The isospectrality between the axial and polar perturbations may not
hold for all the metrics we consider~\cite{Moulin:2019bfh}.
Thus by not including the polar perturbations, the waveform many not be
complete.  
We follow the method used by Gundlach, {\it et
al}~\cite{Gundlach:1993tp}. 

In a finite time domain, the Schr{\"o}dinger-like Eq.~\ref{eq:schr}
can be written as  

\begin{equation}
\left[ \frac{\partial^2}{\partial r_*^2} - \frac{\partial^2}{\partial
t^2} -  V(r_*) \right] \Psi(t,r_*) = 0.
\end{equation}

\noindent
Using light-cone coordinates, $u = t - r_*$ (retarded time) and
$v = t + r_*$ (advanced time), we obtain

\begin{equation}
\left[ 4\frac{\partial^2}{\partial u\partial v} - V(u,v) \right]
\Psi(u,v) = 0.  
\end{equation}

We take the perturbation as a Gaussian wavepacket centered at $v_c =
10$ with a width $\sigma = 3$.
Without loss of generality, we take the initial conditions as 

\begin{equation}
\Psi(0,v) = \exp \left[ -\frac{(v-v_c)^2}{2\sigma^2} \right]  \quad
\textrm{and} \quad \Psi(u,0) = \Psi(0,0).
\end{equation}

\noindent
The observer is placed at future timelike infinity, which we
approximate as $r = 10r_\mathrm{h}$.

The partial differential equation is solved numerically using forward 
differences.
We take $\Delta v = \Delta u = 0.1$, and note that taking $0.05$
provides no visible benefit.
For the purpose of brevity, we choose units in which $r_\mathrm{h} =
1$. 
This means $r, r_*, t$ are expressed in units of $r_\mathrm{h}$.
The QNM frequency $\omega$ is in units of $r_\mathrm{h}^{-1}$ and
units for the QNM potential $V$ are $r_\mathrm{h}^{-2}$.
We have scaled the LQG black hole parameters accordingly.
This is allowed since the wave equation is linear and we are not
interested in the amplitude.

\begin{figure}[p]
\centering
\includegraphics[width=0.49\linewidth]{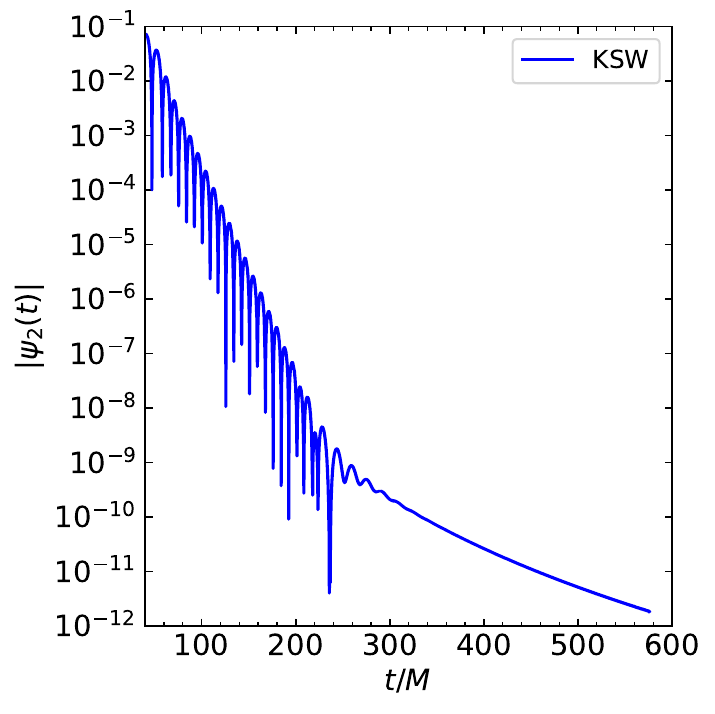}
\includegraphics[width=0.49\linewidth]{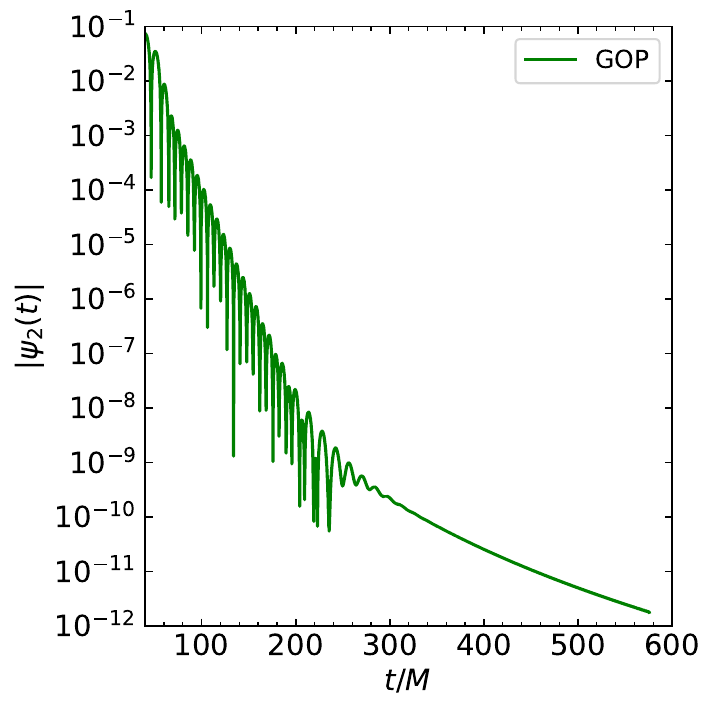}
\includegraphics[width=0.49\linewidth]{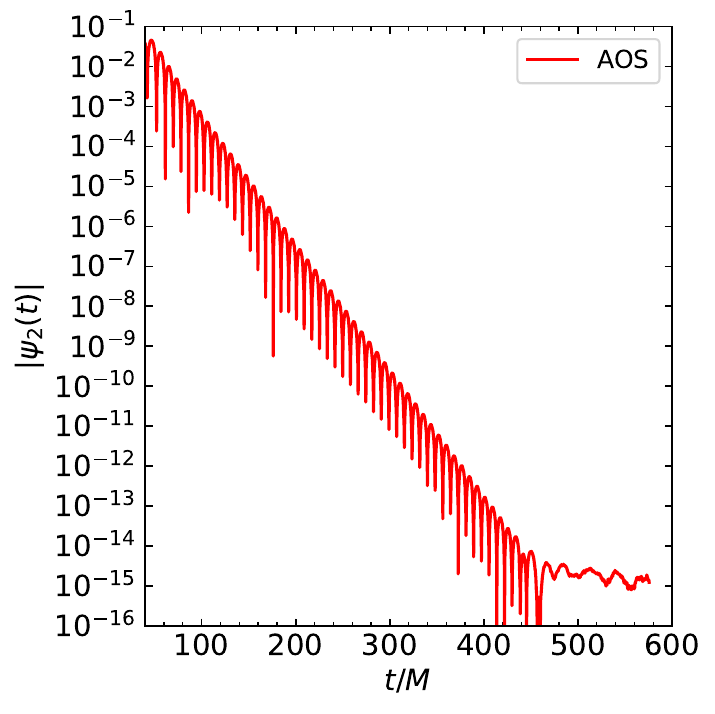}
\includegraphics[width=0.49\linewidth]{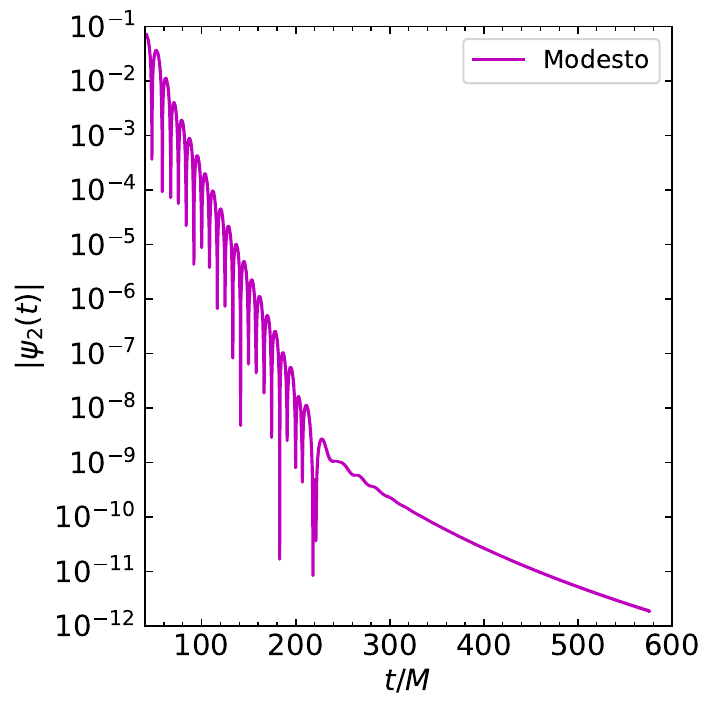}
\caption{\label{fig:ring}(top-left) KSW, (top-right) GOP, (bottom-left)
AOS, (bottom-right) Modesto ringdown waveforms of a Gaussian
wavepacket at $r = 10r_\mathrm{h}$.}
\end{figure}

To validate the procedure, we first consider a Schwarzschild black hole
and obtain the ringdown waveform.
We observe the usual power-law tail at late time.
The result is in agreement with Ref.~\cite{Yang:2023gas}, and others.
Figure~\ref{fig:ring} shows the ringdown waveforms for the LQG black
holes.
The KSW, GOP, and Modesto curves are similar to Schwarzschild case
(not shown). 
The AOS ringdown curve is significantly different.

There are at least three features worth considering in these ringdown
curves.
The time at which the ringing stops is visually the same for all
black holes except AOS which rings considerably longer.
This is due to the significantly different tortoise relationship. 
A second feature is the oscillation frequency.
It is not our purpose in presenting these curves to have an
alternative means of calculating the QNMs. 
However, we simply fit the region $100 \le t/M \le 200$, to the 
function $A e^{\omega_I t} \sin(\omega_R + B)$ and obtain consistent
results to the WKB determination, both for real and imaginary parts of
the frequency; thus partly validating the ringing behavior in the
curves.  
The last feature we will comment on is the power-law tails.
The long-time tails are linear on a log-log plot.
By performing power-law fits in the region $500 \le t/M \le 580$ we
obtain a power of $-7.25$ for all ringdown curves except AOS for which
we obtain $-4.4$, with significantly more uncertainty.
Since our potential is for $s=2$, we can not compare the power-law 
behavior exactly with Ref.~\cite{Gundlach:1993tp} but point out
that we obtain a consistent $1/4$ extra contribution to their 
case of $\ell = 2$ with no initial static field.
The exceptional power-law behavior of AOS is due to the unique
asymptotic behavior at large $r$.
The AOS tail appears less smooth.
As a check, we used the metric approximation of
Ref.~\cite{Daghigh:2020fmw} but obtained the same behavior. 

%
\section{Eikonal limit and circular null geodesics\label{sec:8}}

In this section, we calculate the quasinormal modes in the eikonal, or
geometric optics, limit using the WKB approximation and compare with
the quasinormal modes predicted by circular null geodesics. 
The real part of the complex QNM frequency can be determined by the
angular velocity at the unstable null geodesic and the imaginary part
can be related to the instability time scale of the orbit.
It is important to test these relationships for LQG black
holes~\cite{Konoplya:2017wot} to see how universal they are.
  
Based on the conditions of a stationary, spherically symmetric, and 
asymptotically flat metric, Ref~\cite{Cardoso:2008bp}, and
others, have derived the QNM frequency from the circular null geodesic: 

\begin{equation}
\omega = \left( \ell + \frac{1}{2} \right) \Omega_c - i \left( n +
\frac{1}{2} \right) |\lambda_c|,
\label{eq:omega}
\end{equation}

\noindent
where the subscript $c$ means that the functions are evaluated at the
radius of the circular null geodesic, and $\Omega_c$ and $\lambda_c$
are the coordinate angular velocity and the Lyapunov exponent of the
circular null geodesic, respectively.
The principal Lyapunov exponent is the inverse instability time scale
associated with the geodesic motion.
It is a measure of the average rate at which nearby trajectories
converge or diverge in phase space. 
A positive Lyapunov exponent indicates a divergence between nearby
trajectories.
The WKB results are accurate in the eikonal regime.
Equation~\ref{eq:omega} has been shown to be identical to the
first-order WKB approximation~\cite{Cardoso:2008bp}. 

The radius of the null geodesic $r_c$, is obtained by solving $f_c
h_c^\prime = f_c^\prime h_c$, where the prime means differentiation
with respect to $r$.
The values of $r_c$ can be found in Table~\ref{tab:radii}.
For the Schwarzschild case, $r_c = 3 M$.

The coordinate angular velocity is 

\begin{equation}
\Omega_c = \sqrt{\frac{f_c}{h_c}},
\end{equation}

\noindent
and the positive Lyapunov exponent is

\begin{equation}
\lambda_c = \sqrt{\frac{g_c}{2h_c} (f_c h_c^{\prime\prime} -
  f_c^{\prime\prime} h_c)}\, .
\end{equation}

In the eikonal regime ($\ell \gg 0$), the potentials for all
spin-field perturbations is approximately 

\begin{equation}
V(r) \approx \ell^2 \frac{f}{h}.
\end{equation}

\noindent
Thus the results obtained by solving the Schr{\"o}dinger-like wave
Eq.~\ref{eq:schr} are independent of the spin-field perturbation and
so we only consider the gravitational case.
For GOP, this potential does not depend on $\delta r$.

\begin{figure}[htb]
\centering
\includegraphics[width=0.6\linewidth]{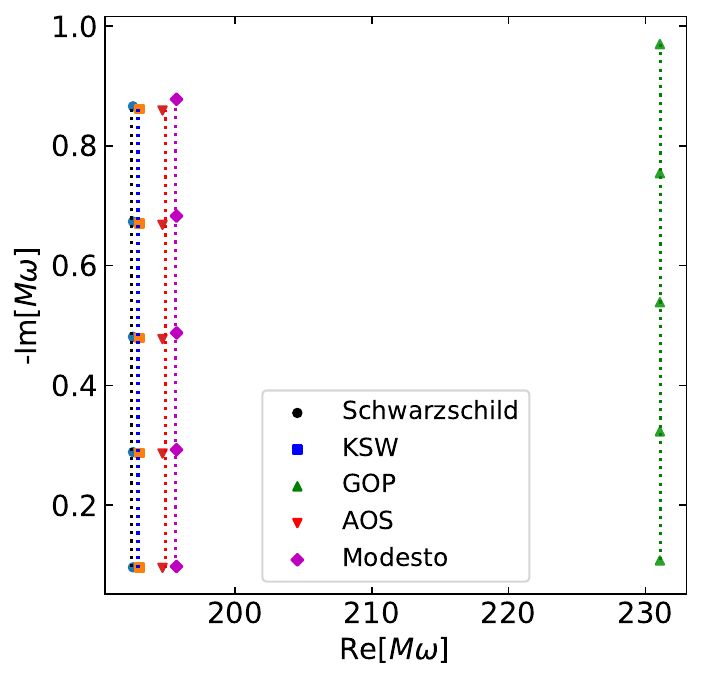}
\caption{\label{fig:eikonal}QNMs of the axial-gravitational
perturbation for $\ell = 1000$ and $n = 0$ to 4 calculated by (solid
markers) the WKB approximation method and (dashed lines) the
QNM circular null geodesic (CNG) relations for $M=3 m_p$.
For CNG, the real frequencies are independent of $n$ and the lines 
between the markers are simply to aid the eye.}  
\end{figure}

Setting $\ell = 1000$ (approximating infinity) and $n = 0$ to $4$, we
calculate the QNM frequencies in the eikonal limit using the WKB
approximation method and using Eq.~\ref{eq:omega}.
The results are shown in Fig.~\ref{fig:eikonal} and the agreement
between the two methods is good for all metrics.
The GOP eikonal results are significantly different due to the smaller
horizon radius and the others are marginally different.

For free, we calculate the shadow radius

\begin{equation}
R_s = \frac{1}{\Omega_c}
\end{equation}

\noindent
of the black holes, which corresponds to the angular semi-diameter of
the shadow around a black hole as seen by an observer at spatial
infinity. 
The results are shown in Table~\ref{tab:radii}.
For the Schwarzschild case, $R_\mathrm{s} = 3\sqrt{3} M$.
The GOP is significantly different due to the smaller circular null
geodesic radius. 

%
\section{Discussion}

We have compared four LQG black hole metrics at $M = 3\Mp$ using
the LQG parameters.
The KSW metric adds a maximum contribution to the metric function
$f(r)$ of 0.8\%.  
KSW appears identical to Schwarzschild in terms of potentials, QNMs,
and ringdown.
Differences of 0.2\% and 0.5\% in the eikonal limit are observed for
the real and imaginary parts, respectively.
To observe the quantum effects of the KSW metric, one needs to
consider black hole masses closer to the Planck scale.

The GOP black hole is the most different compared to the other black
holes.
It is the only metric we examined with an arbitrary free parameter
$\delta r$ which we choose to be the Planck length.
This causes the ADM mass, and hence horizon radius, to be about a
factor of 0.6 small than the other black holes. 
This is a main source of difference.
The validity of the solution forces the ADM mass to be greater than
$2.1\Mp$. 
The QNMs are 20\% and $-12$\% different in the eikonal limit for real
and imaginary parts, respectively.

The main difference for AOS black holes is due to the asymptotic
behavior at large $r$.
However, QNMs are mostly independent of shifts in $r$.
The QNMs in the eikonal limit differ by 1.1\% and 0.8\% in the
real and imaginary part, respectively.
The main difference is in the ringdown.
The ring lasts about 1.6 times longer, with a smaller long-time
power-law slope. 

At $M = 3\Mp$ and using a data-constrained polymeric function value,
the Modesto metric adds a 
maximum contribution to the metric function $f(r)$ of 2.5\% and the
potential is marginally higher at the peak. 
Modesto appears similar to Schwarzschild in terms of QNMs and
ringdown. 
Differences of 1.7\% and $-1.4$\% in the eikonal limit are observed
for the real and imaginary parts, respectively. 
The ability to observe the quantum effects of the Modesto metric is
unlikely, and one needs to consider black hole masses closer to the
Planck scale. 

We have calculate and compare the eigenvalue solutions due to
perturbations on LQG black hole background metrics.
This is the first QNM comparison using common parameters between LQG
black holes.  
Not all possible LQG black hole metrics have been considered, and
others are worthy of investigation. 

The GOP and AOS metrics are deserving of further study and should be
confronted with data.
The parameters of these metrics could be constraint in the way 
the Modesto polymeric function is constrained using data.
Non Schwarzschild like relations between the mass parameter and
horizon radius like in the GOP metric might be constraint by observing
a nominal ringdown but with anomalous QNMs.
Curvature invariants falling off slower than that of Schwarzschild
like in the AOS metirc might be constraint by observing nominal QNMs
but with an anomalously long ringdown time.
Although one should be cautioned against shortfalls in the AOD
metric~\cite{Bouhmadi-Lopez:2019hpp, Bojowald:2019dry,
Ashtekar:2020ckv,  Faraoni:2020stz}. 

Black holes always rotate in the real world, so the gravitational
perturbations of the rotating black holes in loop quantum gravity and
the related properties should be considered in future work.

\section*{Acknowledgments}

I thank Saeed Rastgoo for pointing out some of the LQG
metrics and helpful discussions. 
We acknowledge the support of the Natural Sciences and Engineering
Research Council of Canada (NSERC). 
Nous remercions le Conseil de recherches en sciences naturelles et en
g{\'e}nie du Canada (CRSNG) de son soutien. 
\bibliographystyle{JHEP}
\bibliography{gingrich}
\end{document}